\documentclass[twocolumn,aps,a4paper,showpacs,floatfix,superscriptaddress,nofootinbib]{revtex4}
\usepackage{graphicx,rotating,amsmath,longtable}
\usepackage{lscape}
\usepackage{textcomp}

\begin{document}

\title{Evidence for shape coexistence in $^{98}$Mo}

\author{T.~Thomas\footnote{tim.thomas@ikp.uni-koeln.de}}
\affiliation{Institut f\"ur Kernphysik, Universit\"at zu K\"oln, Z\"ulpicher Stra{\ss}e 77, D-50937 K\"oln, Germany}
\affiliation{Wright Nuclear Structure Laboratory, Yale University, New Haven, Connecticut 06520, USA}
\author{K.~Nomura}
\affiliation{Grand Acc\'el\'erateur National d'Ions Lourds, CEA/DSM-CNRS/IN2P3, Boulevard Henri Becquerel, F-14076 Caen Cedex 05, France}
\affiliation{Institut f\"ur Kernphysik, Universit\"at zu K\"oln,
Z\"ulpicher Stra{\ss}e 77, D-50937 K\"oln, Germany}
\author{V.~Werner}
\affiliation{Wright Nuclear Structure Laboratory, Yale University, New Haven, Connecticut 06520, USA}
\author{T.~Ahn}
\affiliation{Wright Nuclear Structure Laboratory, Yale University, New Haven, Connecticut 06520, USA}
\author{N.~Cooper}
\affiliation{Wright Nuclear Structure Laboratory, Yale University, New Haven, Connecticut 06520, USA}
\author{H.~Duckwitz}
\affiliation{Institut f\"ur Kernphysik, Universit\"at zu K\"oln, Z\"ulpicher Stra{\ss}e 77, D-50937 K\"oln, Germany}
\author{M.~Hinton}
\affiliation{Wright Nuclear Structure Laboratory, Yale University, New Haven, Connecticut 06520, USA}
\affiliation{Department of Physics, University of Surrey, Guildford GU2 7XH, United Kingdom}
\author{G.~Ilie}
\affiliation{Wright Nuclear Structure Laboratory, Yale University, New Haven, Connecticut 06520, USA}
\author{J.~Jolie}
\affiliation{Institut f\"ur Kernphysik, Universit\"at zu K\"oln, Z\"ulpicher Stra{\ss}e 77, D-50937 K\"oln, Germany}
\author{P.~Petkov}
\affiliation{Institut f\"ur Kernphysik, Universit\"at zu K\"oln, Z\"ulpicher Stra{\ss}e 77, D-50937 K\"oln, Germany}
\affiliation{Bulgarian Academy of Science, Institute for Nuclear Research and Nuclear Energy, Tsarigradsko Chausse 72, 1784 Sofia, Bulgaria}
\author{D.~Radeck}
\affiliation{Institut f\"ur Kernphysik, Universit\"at zu K\"oln, Z\"ulpicher Stra{\ss}e 77, D-50937 K\"oln, Germany}

\date{\today}

\begin{abstract}
A $\gamma\gamma$ angular correlation experiment has been performed to
 investigate the low-energy states of the nucleus $^{98}$Mo. The new
 data, including spin assignments, multipole mixing ratios and lifetimes
 reveal evidence for shape coexistence and mixing in $^{98}$Mo, arising
 from a proton intruder configuration. This result is reproduced by a
 theoretical calculation within the proton-neutron interacting boson
 model with configuration mixing, based on microscopic energy density
 functional theory. The microscopic calculation indicates the importance
 of the proton particle-hole excitation across the $Z=40$ sub-shell
 closure and the subsequent mixing between spherical vibrational and the
 $\gamma$-soft equilibrium shapes in $^{98}$Mo.  
\end{abstract}

\maketitle

For decades, to clarify the nature of shape coexistence 
has been one of the major objectives in nuclear structure 
physics \cite{wood92,Heyde11}. 
The phenomenon has been observed in various regions of the nuclear
chart, from light \cite{Bro66a} 
to heavy \cite{Andr00} systems. In $^{186}$Pb, for example, three
low-lying 0$^+$ states bunch together in energy, within the range of 700
keV \cite{Andr00}. The emergence of the extremely low-lying $0^{+}$
states is, 
in terms of the spherical shell model, attributed to two- or four- proton
excitations across the $Z=82$ shell closure. The 
residual interaction between protons and neutrons leads to the lowering
of the excited $0^{+}$ states and the different corresponding shell-model configurations
are linked to relevant geometrical deformations in a mean-field picture
\cite{naza93}.

The $A\sim 100$ mass region also presents a unique laboratory for the
evolution of nuclear shape and shape coexistence
\cite{VW02,Simp06}. The interplay between single-particle and collective
degrees of freedom leads to shape phase transitions along isotopic and
isotonic chains \cite{cejnar10rev}. The most dramatic examples for shape
coexistence and shape transition occur in the Zr isotopic
chain, as recently revealed for $^{94}$Zr~\cite{Chak13}. 
Especially in the $N=50-56$ Zr isotopes the $0^+_1$ state and the
very low-lying $0^+_2$ state are considered strongly mixed $0p$-$0h$ and $2p$-$2h$
proton configurations, where protons are promoted from the $pf$ shell to
the $g_{9/2}$ orbital, as also found in shell model
calculations~\cite{VW02, Sieja09}. 
The structure of the low-lying
$0^+_2$ state in $N \ge 58$ Zr isotopes is somewhat more complicated due
to neutron contributions. 
In Mo isotopes, starting from $N=50$ the nuclear shape
gradually evolves from a sphere and, driven by the enhanced
proton-neutron residual interaction, large deformation sets in at
$N\approx 60$~\cite{Fed79}. Situated in between, $^{98}_{42}$Mo$_{56}$
is pivotal for understanding shape transitions in this mass region. In
particular, the concept of shape coexistence can apply to this nucleus,
where proton cross-shell excitations from the $Z=28-40$ $pf$ shell to the
$\pi g_{9/2}$ orbit may play an important role~\cite{Rusev05}. 
In fact, experimentally, the first excited state of $^{98}$Mo has been shown to
be an coexisting isomeric 0$^+$ state of different shape~\cite{Bur72,Ziel02}. The mixing between the proton $2p$-$0h$ and $4p$-$2h$ configurations
forms the first excited 0$^+$ state and the ground state as revealed by
the investigation of $\gamma$ transitions depopulating 1$^+$ states with
equal strengths to both $0^{+}$ states~\cite{Rusev05}, akin to the
findings for $^{92}$Zr~\cite{VW02}.

To address the important issue of the nature of low-lying structure in $^{98}$Mo, 
we performed a $\gamma \gamma$ angular correlation experiment. 
In this paper, the results of this experiment are reported as well as the
identification of shape coexistence and the role of a proton intruder
configuration in $^{98}$Mo. 
The experimental results are supported by predictions of the
interacting boson model \cite{IBM} with configuration mixing, where
the Hamiltonian is determined microscopically. 
The microscopic calculation indicates the importance of the proton
intruder configuration and the substantial mixing between spherical-vibrational 
and $\gamma$-unstable shapes in $^{98}$Mo.

In order to extend the $^{98}$Mo level scheme, we used the
reaction $^{96}$Zr($\alpha$,2n)$^{98}$Mo. A 16 MeV $\alpha$ beam was delivered by the ESTU tandem
accelerator at the Wright Nuclear Structure Laboratory, Yale University, impinging on a 1.25
mg/cm$^2$ thick $^{96}$Zr target enriched to 57.36$\%$. The $\gamma$~transitions were
detected by 10 Compton-suppressed HPGe Clover detectors of the YRAST
Ball array~\cite{beau00}. During five days of measurement, 1.2$\times
10^{9}$ events were collected using a $\gamma \gamma$ coincidence
trigger.

\begin{figure}[!ctb]
   \centering
        \includegraphics[scale=0.38]{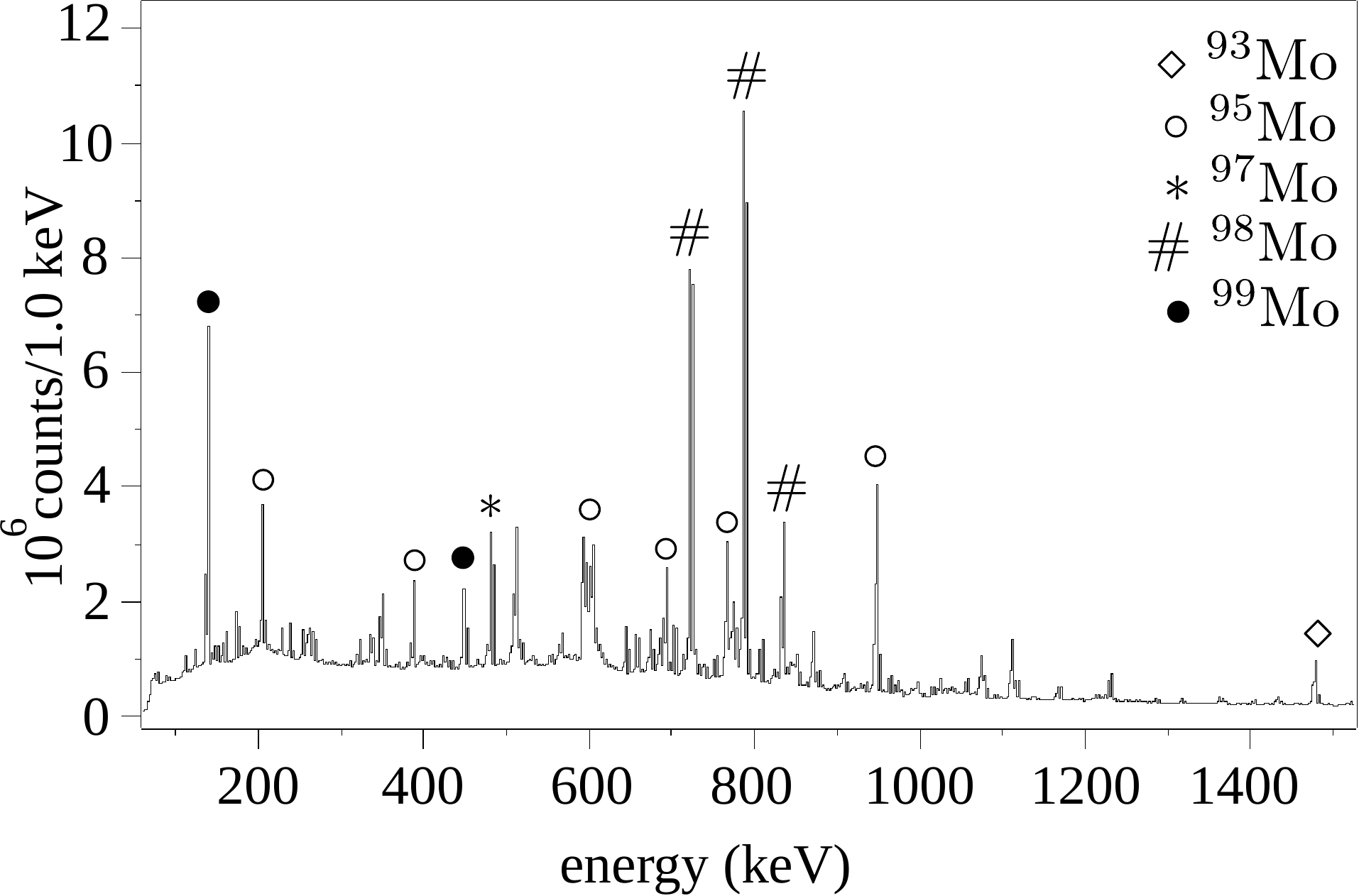}
         \caption{\label{gammaspectrum} Total projection of the $\gamma \gamma$ coincidence data. Major peaks from $^{98}$Mo and the main side reactions are marked.}
\end{figure}

\begin{figure}[!ctb]
   \centering
        \includegraphics[scale=0.55]{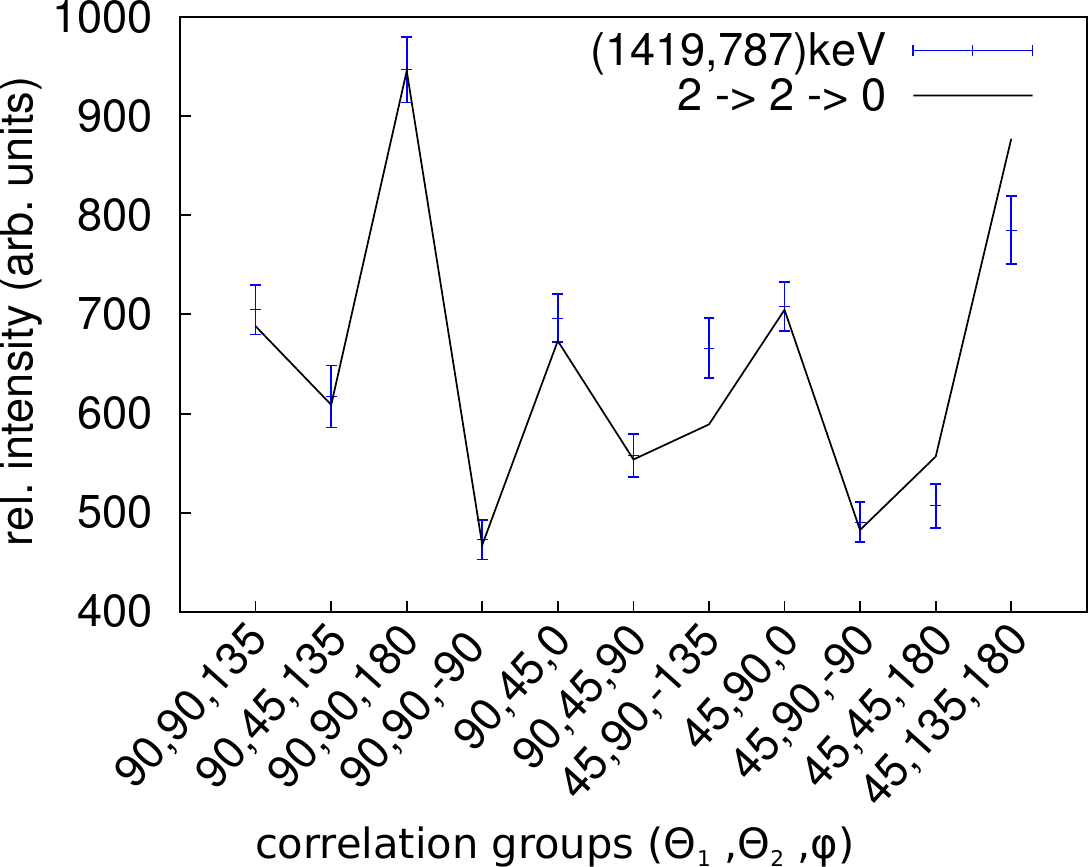}
         \caption{\label{correlation} (Color online) Comparison of a fitted theoretical angular correlation (solid line) with relative  intensities obtained from 11 correlation groups for the 1419-787 keV $\gamma \gamma$ coincidence.}
\end{figure}

Figure \ref{gammaspectrum} shows the total projection of the $\gamma
\gamma$ coincidence data. Due to impurities in the $^{96}$Zr target
transitions from $^{93-99}$Mo isotopes were observed. The most prominent
peaks are labeled with their associated nuclear origin. The data was
sorted into 11 correlation group matrices, which account for detector
pairs at angles $\Theta_{1,2}$ with respect to the beam axis and a
relative angle $\psi$ between the plains spanned by the detectors and the
beam axis, in order to perform a $\gamma \gamma$ angular correlation
analysis. Relative intensities in
the correlation groups were then fitted to angular correlation functions to 
extract spins and multipole mixing ratios, as described in Refs.~\cite{kra70,kra73}, using the computer 
code CORLEONE~\cite{corleone,corleoneII}. The code takes into
account the attenuation factors of the detectors~\cite{Ros53,Casp09}. An
example of a $\gamma \gamma$ angular correlations analysis is shown in
Fig.~\ref{correlation} for the 2$^+_4$ $\xrightarrow{1419}$ 2$^+_1$
$\xrightarrow{787}$ 0$^+_{gs}$ cascade, yielding the hitherto unknown
multipole mixing ratio $\delta_{1419}=0.33\pm 0.11$. In
literature~\cite{NDS03} conflicting multipole mixing ratios are given
for $\gamma$ transitions depopulating low-lying states in $^{98}$Mo. The
superior sensitivity of the present setup allowed to resolve
discrepancies. For more detailed information about $\gamma \gamma$
angular correlations analysis with the YRAST Ball array see
Refs.~\cite{Casp09,Will09}.  
In the same way, the multipole mixing ratio of the 2$^+_2$ $\xrightarrow{644}$ 2$^+_{1}$ transition
was measured to be +1.67(25), which is in agreement with the larger
solution from an (n,n$^{\prime}\gamma$) experiment~\cite{Mey84}, and refutes the most recent
value from Coulomb excitation~\cite{Ziel02}.
%--------------------------------------------------------------
%--------------------------------------------------------------

%--------------------------------------------------------------
%--------------------------------------------------------------

%--------------------------------------------------------------
\begin{figure}[!ctb]
   \centering
        \includegraphics[scale=0.51]{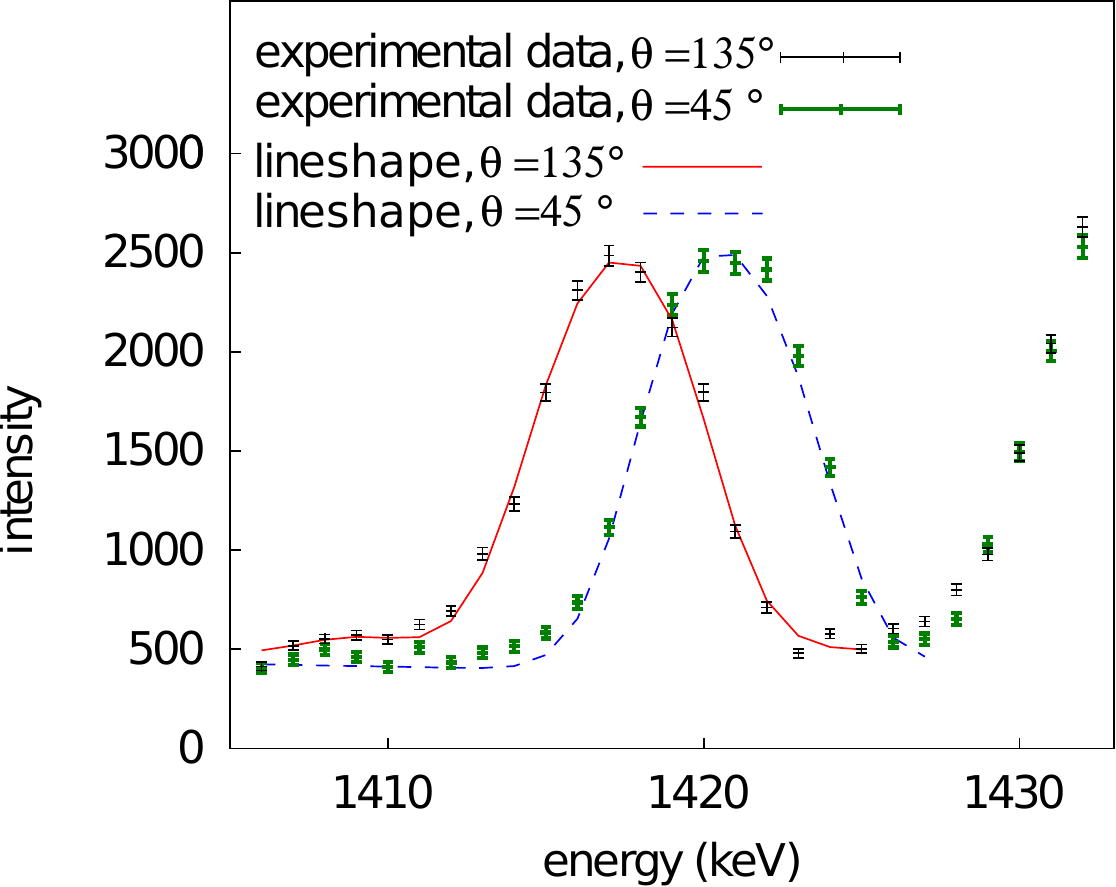}
         \caption{\label{dsam} (Color online) Determination of the effective lifetime of the 1419 keV transition depopulating the 2$^+_4 $ state using a gate set on the 787 keV transition. 
         Coincidence spectra with a gate set on the 787 keV transition for two different angles are shown. The red solid line represents the simulated line shape at forward angle and 
         the blue dashed line the backward angle. The effective average lifetime is $\tau$=0.30 (7) ps.}
\end{figure}
%--------------------------------------------------------------

Lifetimes of excited states were determined using the
\textit{Doppler-Shift Attenuation Method} (DSAM)~\cite{Cur68}. 
%The
%recoiling nuclei may emit $\gamma$ radiation during the stopping process
%in the target material. Therefore, some of the $\gamma$ radiation is
%detected with a Doppler shift in the forward or backward direction. 
The data was sorted into three matrices according to the three angles
$\theta$=45$^\circ$, 90$^\circ$, 135$^\circ$ of the detectors relative
to the beam axis. 
%The lifetime of a state is obtained by reproducing the
%line shape of a partially Doppler shifted peak. 
For the line shape analysis, the stopping 
process of an excited nucleus is simulated using
nuclear~\cite{North70} and electronic stopping powers~\cite{Zieg78}. In
Fig. \ref{dsam}, a line shape analysis for the 1419 keV transition
depopulating the 2$^+_4$ state is shown. The weighted mean value over
the angles for the effective lifetime is calculated to be $\tau$=0.30 (7) ps. The analysis procedure is outlined in more detail
in Ref.~\cite{Pet98}. %The complete results obtained from this in-beam
		      %experiment for $^{98}$Mo will be published in a
		      %subsequent paper \cite{Thom13}. 

\begin{figure}[!ctb]
   \centering
        \includegraphics[scale=0.21]{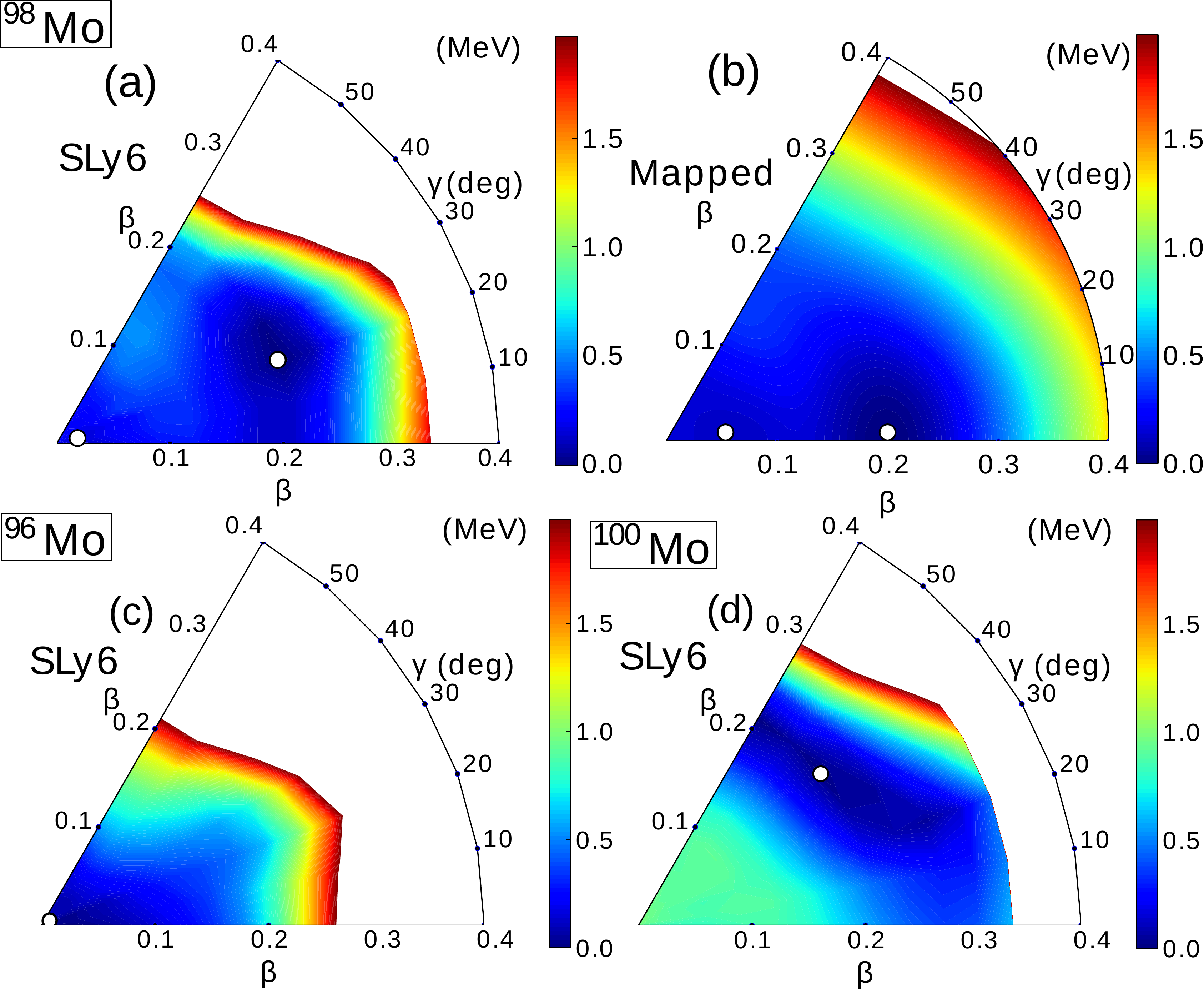}
         \caption{(Color online) Contour plots
 of the microscopic (a) and the mapped (b) energy surfaces in ($\beta$,
 $\gamma$) plane of $^{98}$Mo, and of the microscopic energy surfaces of
 the adjacent nuclei $^{96}$Mo (c)
 and $^{100}$Mo (d). The color code ranges from 0 (mean-field minimum)
 to 2 MeV, and the minima are identified by the solid white circles. The Skyrme SLy6 functional is used. 
} 
\label{fig:pes}
\end{figure}

To interpret the nature of the low-lying structure and the relevant
shape dynamics in $^{98}$Mo, we performed a self-consistent mean-field
calculation using the Skyrme energy density functional (EDF) (see~\cite{ben03rev} for review). 
Figure~\ref{fig:pes}(a) shows 
the total energy surface of $^{98}$Mo in 
terms of the axial quadrupole deformation $\beta$ and triaxiality
$\gamma$~\cite{BM}, obtained through the constrained Hartree-Fock-BCS
(HF-BCS) method with the Skyrme functional SLy6 \cite{Chab98} using the code 
ev8 \cite{Bonche05}. 
%The energy surface refers to the total mean-field energy as a function
%of $\beta$ and $\gamma$, obtained as a solution of the
%HF-BCS calculation, and symmetry projection is not performed in the present work. 
Figure~\ref{fig:pes}(a) displays two minima in the mean-field
energy surface, with the deeper one being close to a spherical
shape ($\beta \approx 0$) and the other at $\beta \approx 0.21$ and
$\gamma\approx 20^{\circ}$ with some degree of softness. 
On the other hand, no coexisting minima are visible in the microscopic
energy surfaces of the adjacent 
nuclei $^{96}$Mo (Fig.~\ref{fig:pes}(c)) and $^{100}$Mo
(Fig.~\ref{fig:pes}(d)). 
$^{98}$Mo appears to be transitional between near-spherical ($^{96}$Mo)
and deformed ($^{100}$Mo) shapes.

%-----------------------------------------------

To study quantitatively the spectroscopic observables associated with the intrinsic
shape of interest, it is necessary to go beyond the mean-field
approximation. 
In this work we resort to the proton-neutron
interacting boson model (IBM-2)~\cite{IBM,OAI} to generate spectra and
transition rates that are comparable to data. 
%The Hamiltonian for the 
%IBM-2 is determined, following the method of \cite{Nom08}, by mapping
%the microscopic energy surface onto the expectation value of the
%IBM-2 Hamiltonian in the boson condensate \cite{GK}. 
By mapping the microscopic energy surface onto the equivalent
IBM-2 Hamiltonian in the boson condensate~\cite{GK}, 
the Hamiltonian parameters are determined microscopically, 
thereby not invoking any adjustment to data (cf. Ref.~\cite{Nom08,Nom12sc} for details). 
The mapped Hamiltonian is to be diagonalized numerically in the boson $m$-scheme
basis to provide level energies and transition rates with good quantum
numbers in the laboratory frame. 

In order to describe the two mean-field minima, the model space of the IBM-2 needs to be
extended by including the intruder configuration 
and by mixing the Hamiltonians associated with the two configurations
\cite{duval81}. 
We include the two-proton excitation across the $Z=40$ shell,
assuming $^{90}$Zr to be the inert core. 
Thus, the number of proton bosons is 1 
and 3 for the normal and the intruder configurations, respectively,
while the neutron boson number is fixed to 3. 
Note that normal (intruder) configuration denotes hereafter the proton $2p$-$0h$
($4p$-$2h$) configuration. 
The choice of this model space is also guided by studying the calculated single-particle
energies as functions of the $\beta$ deformation, which indicate the
lowering of $\pi g_{9/2}$ orbitals and the occupation of the last protons in the
orbitals at $\beta\approx 0.2$ associated to the $\gamma$-soft minimum in Fig.~\ref{fig:pes}(a). 
The full Hamiltonian of the system is then given as: 
\begin{eqnarray}
 H=P_{nor}H_{nor}P_{nor}+P_{intr}(H_{intr}+\Delta)P_{intr}+H_{mix},
\end{eqnarray}
where $H_{nor}$ ($H_{intr}$) and $P_{nor}$ ($P_{intr}$) represent the
Hamiltonian of and the projection operator onto the normal (intruder)
configuration space, respectively. 
$\Delta$ and $H_{mix}$ stand for the energy 
offset needed for the proton cross-shell excitation and interaction that
mixes two configurations, respectively. 
The form of the Hamiltonian $H$ is the same as in \cite{Nom12sc}, but only two configurations are
considered in the present work. 
%The parameters of the two independent IBM-2 Hamiltonians, the mixing strength, and the
%energy offset are determined solely from the microscopic energy surface,
The resulting mapped IBM-2 energy surface is
shown in Fig.~\ref{fig:pes}(b). One can see in Fig.~\ref{fig:pes}(b) two
equivalent minima near $\beta \approx 0$ and $\beta \approx 0.2$, with
the latter being $\gamma$ soft similarly to the microscopic energy
surface~\footnote{A minimum at $\gamma=20^{\circ}$, however, is not obtained
with the used Hamiltonian containing up to only two-body boson terms. It has been shown \cite{Nom12tri} that a three-body boson term
should be included in the IBM Hamiltonian to give rise to the triaxial
minimum and to better describe the detailed structure of quasi-$\gamma$
band. This is, however, not particularly of relevance for the present paper.}. 

The calculation predicts a spectroscopic quadrupole moment for the 
$2^+_1$ state of $Q(2^+_1) = -0.245$ $e$b, corresponding to a weak prolate deformation.
This is consistent with a previous experimental value of $Q(2^+_1) = -0.25(9)$
$e$b~\cite{Pa79}, but differs from the more recent one, $Q(2^+_1)
=-0.05(2)$ $e$b \cite{Ziel02}. 
We note, that the latter result stems from a global fit to data taking
known multipole mixing ratios and lifetimes into account. 
Some of these input data have been changed and complemented by our present
measurement. In Table I, we give the intrinsic $\beta$-deformation
parameters for the lowest three $2^+$ states, taken from inelastic
scattering~\cite{Pe77,Uk01} and Coulomb excitation~\cite{Ba72} data. This data is
compared to the value obtained from the minima in the mean field energy
surface (Fig. 4(a)), and the deformation extracted from the intrinsic
quadrupole moment in the IBM-2, assuming $K=0$. The best agreement is
found with Coulomb excitation values from Ref.~\cite{Ba72}. 
\begin{table}[!cb]
 \caption{\label{beta_2} The intrinsic deformation parameter $\beta_{2}$
 for the lowest three excited $2^{+}$ states. 
The theoretical values extracted from the intrinsic quadrupole moments
 obtained by the IBM-2 ($K=0$ is assumed) $\beta_{2}^{\textnormal{IBM}}$, and the equivalent
 values $\beta_{2}^{\textnormal{MF}}$ associated with the mean-field
 minima, and the experimental values $\beta_{2}^{\textnormal{expt}}$
 from inelastic scattering of deuterons~\cite{Uk01,Pe77} and Coulomb
 excitation~\cite{Ba72} are shown.}
\begin{tabular}{cccccc}
\hline
\hline
$E_{\textnormal{level}}$ (keV) & $J^{\pi}$ &
 $\beta_{2}^{\textnormal{MF}}$ &
 $\beta_{2}^{\textnormal{IBM}}$ &  $|\beta_{2}^{\textnormal{(d,d')}}|$   & $|\beta_{2}^{\textnormal{CoulEx}}|$ \footnotemark[1]  \\ \hline
787.26  & 2$^+_1$ & (+0.21) & +0.132  & 0.167~(4) \footnotemark[2]  & 0.174~(5) \\ 
1432.29 & 2$^+_2$ & ($\approx $0.0) & +0.060 & 0.046 \footnotemark[3]  & 0.037~(2) \\
1758.32 & 2$^+_3$ & & -0.121  & 0.029 \footnotemark[3] & 0.11~(5)     \\
\hline
\hline
\end{tabular}
\footnotetext[1]{Taken from Ref.~\cite{Ba72}} \footnotetext[2]{Taken from Ref.~\cite{Uk01}}
\footnotetext[3]{Taken from Ref.~\cite{Pe77}}
\end{table}

%---------------------------------------------------------------------------------------
\begin{figure*}[!ctb]
   \centering
        \includegraphics[scale=0.27]{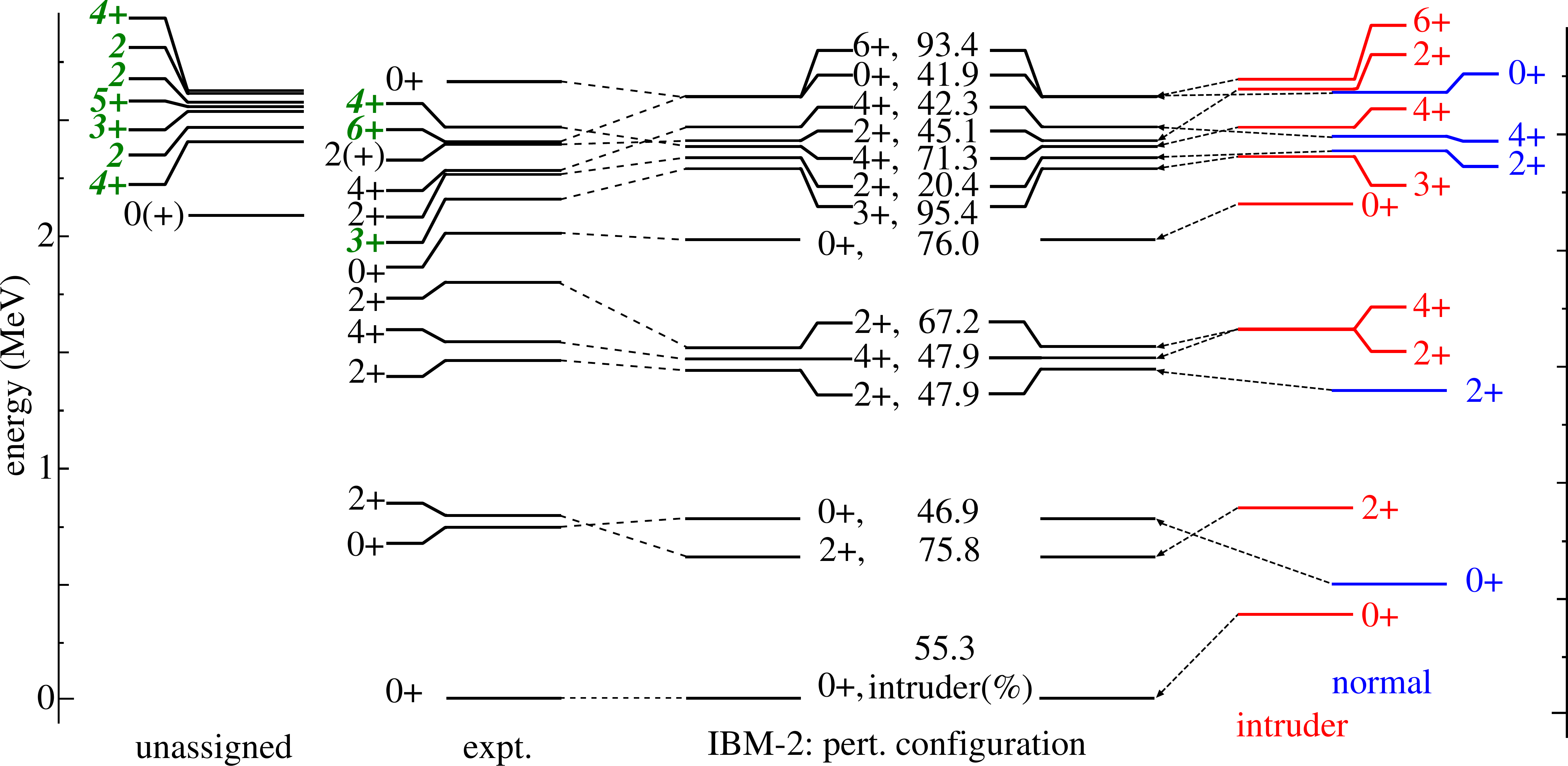}
         \caption{\label{levelscheme} (Color online) Low-energy level
 scheme of $^{98}$Mo. The experimental (left) and the calculated spectra
 with mixing (``IBM-2: pert. configuration'', center) and without mixing
 (right). New spin assignments are
denoted in italic letter. The number indicated next to spin value (center) represents the
 fraction of the intruder configuration in the wave function of each
 state. }
\end{figure*}
%---------------------------------------------------------------------------------------

Next we analyze the structure of the low-energy level scheme of
$^{98}$Mo. Figure \ref{levelscheme} compares  the data from the present
experiment (left-hand side) and the calculated spectra after (center)
and before the mixing, i.e., unperturbed configurations (right-hand
side). Note that some experimental states, which are close in energy and
have the same spin have been identified from the comparison to predicted
$B$(E2) values (cf. Tables \ref{transstr} and \ref{reltrans}). 
Even though the energy levels are 
calculated without any fit to data, that is, the Hamiltonian parameters
are derived solely from the microscopic EDF and the mapping procedure,
the overall agreement between data and calculation in
Fig.~\ref{levelscheme} 
is remarkably good. 
While the
experimental $0^{+}_{2}$ excitation energy is well reproduced 
by the theory, the calculated $2^{+}_{1}$ level energy  seems rather low
compared to the experimental value.  
The reason is the strong level repulsion between the unperturbed
low-spin states of the two configurations due to a rather large mixing
strength.  
In the experiment an excess of states is observed above the $0^+_4$
state, which could not be assigned to predicted states. These might
originate from a more complicated structure eventually associated to
higher-order effects such as the four-proton cross-shell excitation and/or
the excitation of neutrons, which are outside of the model space of the present
calculation.  

Looking into the origin of each state in a more quantitative manner,
firstly we notice on the 
right-hand side of Fig.~\ref{levelscheme} that, the
unperturbed $0^{+}_{1,2}$ states of the normal and the intruder 
configurations are very close in energy. After the mixing, the $0^{+}$ ground
states in each configuration repel each other by $\approx$350 keV in
energy (as illustrated by arrows). 
Here, the matrix element $\langle
H_{mix}\rangle$, which mixes unperturbed $0^{+}_{1}$ states of the
normal and the intruder configurations, is calculated to be 385 keV. This value is consistent
with the result from a schematic two-level mixing calculation of 326 keV
\cite{Rusev05}.  
In the resulting $0^{+}_{1,2}$ states, normal
and intruder configurations are almost equally
mixed with fraction of 55.3\% and 46.9\%, respectively.  

One should also notice that the unperturbed normal and the intruder
level schemes exhibit typical vibrational and $\gamma$-soft
characteristics, respectively which is at variance with the findings in Ref.~\cite{Ziel02}. 
Within the unperturbed normal configuration, the level-spacing with an
$R_{4/2}$ value of 2.67, and the absence of an excited $0^+$ state near the
close-lying $4^+_{1}$, $2^+_{2}$ states, is typical for a gamma-soft structure
($R_{4/2}\sim 2.32$), in which a two-phonon $0^+$ state is absent. The unperturbed
intruder configuration, in contrast, displays close-lying $4^+_{1}$, $2^+_{2}$,
and $0^+_2$ states, more typical for a spherical vibrator. Also the $R_{4/2}=2.32$ 
of the intruder configuration deviates strongly from
deformed values toward the spherical harmonic oscillator ($R_{4/2}= 2.0)$.
This interpretation correlates
with the microscopic energy surface in  Fig.~\ref{fig:pes}(a), and is
consistent with previous empirical IBM-2 fitting calculations
\cite{Samb82}.  

%For the normal configuration, $R_{4/2}=2.32$ is near the vibrational or U(5)
%limit ($=2.0$) of the IBM \cite{IBM} and the ($0^{+}_{2}$, $2^{+}_{2}$,
%$4^{+}_{1}$) multiplet is seen. 
%In the intruder scheme, the 
%$R_{4/2}=2.67$ is close to the $\gamma$-unstable or O(6) limit ($=2.5$)
%of the IBM \cite{IBM}, and one sees that the $2^{+}_{2}$ and
%the $4^{+}_{1}$ states form a doublet. Therefore the low-lying structure
%of $^{98}$Mo can be traced back to the coupling between spherical
%vibrational and the $\gamma$-soft shapes. This interpretation correlates
%with the microscopic energy surface in  Fig.~\ref{fig:pes}(a), and is
%consistent with previous empirical IBM-2 fitting calculations
%\cite{Samb82}.  

\squeezetable
\begin{table}[!h]
 \caption{\label{transstr} Theoretical E2 transition strengths (in W.u.)
 compared
 to experimental values from \cite{NDS03, Ziel02} and from this
 work. States in bold are predicted to be of intruder nature in theory. For transitions with mixed multipolarity the multipole mixing ratio $\delta$ measured in the present experiment is given.} 
\begin{tabular}{ccccccc}
\hline
\hline
$E_{\textnormal{level}}$ (keV) & $J^{\pi}_I$ & $E_\gamma$ (keV) & $J^{\pi}_F$  & $B({\textnormal{E2}})_{\textnormal{theo}}$ & $B({\textnormal{E2}})_{\textnormal{exp}}$ & $\delta_{expt}$ \\
 \hline
787.26\footnotemark[1]  & {\boldmath $2^+_1$} & 787.26  & {\boldmath $0^+_1$}              & 27 & 21.4 $^{+11}_{-10}$& \\ 
        &         & 52.6\footnotemark[2]   & 0$^+_2$           & 256  & 280 (40) &   \\
1432.29\footnotemark[1] & 2$^+_2$ & 644.70  &  {\boldmath $2^+_1$}   & 22 & 47.8$^{+132}_{-100}$ & $+1.67$ (25) \\
        &         & 697.10  & 0$^+_2$       & 8  & 2.5$^{+8}_{-6}$ &  \\
        &         & 1432.29 & {\boldmath $0^+_1$}             & 0.03  & 1.0$^{+2}_{-1}$ & \\
1509.74\footnotemark[1] & 4$^+_1$ & 722.48  &  {\boldmath $2^+_1$}      & 49 & 49.1$^{+5.5}_{-4.5}$&  \\
1758.32\footnotemark[1] & {\boldmath $2^+_3$} & 326.05  &  2$^+_2$             & 13 & 4.7$^{+189}_{-23}$ & $-0.17$ (22) \\
        &         & 971.03  & {\boldmath $2^+_1$}       & 6   & 3.2$^{+134}_{-16}$ & $-0.97$ (14) \\
        &         & 1023.61 & 0$^+_2$               & 7 & 7.8$^{+286}_{-34}$ & \\
2206.74 & 2$^+_4$ & 1419.48 &  {\boldmath $2^+_1$}       &  1.3 & 1.7 (2) &  $-0.33$ (11) \\
2333.03 & 2$^{(+)}_5$ &900.85 &  2$^+_2$   &  1 & 1.6$^{+8}_{-4}$ &$-0.15^{+0.19}_{-0.20}$ \\
2343.26\footnotemark[3] & {\boldmath $6^+_1$} &  833.52 &  4$^+_1$          & 56 & 10.1 (4)  &  \\
\hline
\hline
\end{tabular}
\footnotetext[1]{$\tau$ adopted from~\cite{Ba72}}
\footnotetext[2]{$B$(E2) adopted from~\cite{NDS03}}
\footnotetext[3]{$\tau$ adopted from~\cite{Ziel02}}
\end{table}

Finally, in Tables \ref{transstr} and \ref{reltrans} we compare
experimental and theoretical $B$(E2) values. Lifetimes are either
adopted from Ref. \cite{Ziel02} or measured in the present
experiment. If not stated differently, all multipole mixing ratios and branching ratios are from 
the present work. The conversion coefficient $\alpha$ was obtained from calculations using the code  {\em Bricc}~\cite{bricc}. Very good agreement between experiment
and theory is obtained, confirming the strong mixing between both configurations. 
In particular, the strong $B({\textnormal{E2}};2^+_1 \rightarrow 0^+_2)$ and
$B({\textnormal{E2}};2^+_2 \rightarrow 2^+_1)$ transitions ,relative to
the $2^+_1 \rightarrow 0^+_1$ transition (see Table \ref{transstr}), present a stringent test
of configuration mixing. The measured $B({\textnormal{E2}};6^+_1
\rightarrow 4^+_1$) is much smaller than predicted, perhaps due to
fragmentation. 

\squeezetable
\begin{table}[!ht]
\caption{\label{reltrans} Same as Table \ref{transstr}, but normalized
 with respect to the largest $B$(E2) value among the depopulating decays
 from a given initial state. }
\begin{tabular}{ccccccc}
\hline
\hline
$E_{\textnormal{level}}$ (keV) & $J^{\pi}_I$ & $E_\gamma$ (keV) & $J^{\pi}_F$ &
 $B({\textnormal{E2}})^{\textnormal{rel}}_{\textnormal{theo}}$ &
 $B({\textnormal{E2}})^{\textnormal{rel}}_{\textnormal{expt}}$ & $\delta_{expt}$\\
\hline
1962.81 & {\boldmath $0^+_3$} 	& 530.61 & 2$^+_2$ & 1 & 1 &  \\ 
 & 			& 1175.57& {\boldmath $2^+_1$} & 0.10 & 0.05 (1) &   \\ 
2104.66 & {\boldmath $3^+_1$} 	& 594.65 \footnote[1]{Branching ratio adopted from~\cite{NDS03}, no multipole mixing ratio available, assumed to be a pure E2 transition} & 4$^+_1$ & 0.66 & $<$ 0.40  &   \\
 & 		& 672.50 & 2$^+_2$& 1 & 1 & +6.66$^{+3.41}_{-1.71}$ \\  
 & 		& 1317.37& {\boldmath $2^+_1$} & 0.13 & 0.04 (3) & $+2.91^{+0.64}_{-0.46}$ \\ 
2223.74 & {\boldmath $4^+_2$}  & 713.80 & 4$^+_1$  & 1 & 1 & $+1.13$ (17) \\
 & 		& 791.58 & 2$^+_2$& 1.60 & 0.88 (11)& \\  
 & 		& 1436.68& {\boldmath $2^+_1$} & 0.03 & 0.04 (1)&  \\ 
2419.48 & {\boldmath $4^+_4$}  & 661.16 & {\boldmath $2^+_3$} & 1 & 1&  \\
 & 		& 909.52 & 4$^+_1$& 0.54 & 0.33 (3)& $-0.64$(10)\\  
 & 		& 1632.46& {\boldmath $2^+_1$} & 0.06 & 0.02 (1) & \\ 
\hline
\hline
\end{tabular}
\end{table}

In Table \ref{reltrans} we compare relative $B$(E2)
values, normalized with respect to the largest $B$(E2) value among the
depopulating decays from a given initial state, for the states without
lifetime information.  
Note, that the three 4$^+_{2,3,4,exp}$ states are observed within 200
keV. From comparison of relative $B$(E2) values the 4$^+_{2,exp}$ state
can be assigned to the predicted 4$^+_{3,theo}$ state generated mainly
by the intruder configuration, while the 4$^+_{4,exp}$ state can be
assigned to a strongly mixed 4$^+_{2,theo}$ state.  Table \ref{reltrans}
shows the same extent of consistency as obtained in Table
\ref{transstr}.

We have revealed robust experimental evidence for shape coexistence and
 configuration mixing in the low-lying structure of $^{98}$Mo. 
Key data on multipole
 mixing ratios and lifetimes have been obtained, allowing for a detailed
 comparison with a new theoretical calculation within
 the IBM based on the microscopic EDF. 
The EDF calculation predicted two (near-spherical and $\gamma$-soft) mean-field minima in the energy 
 surface (Fig.~\ref{fig:pes}(a)), which necessitates the extension of
 the IBM to include a intruder configuration associated to the proton excitation across the
 $Z=40$ subsell closure. 
The two intrinsic shapes are mixed strongly into low-spin states (cf. Fig.~\ref{levelscheme}). 
The excitation spectra and E2 properties are calculated in a fully microscopic way, and are in excellent
 agreement with the wealth of the new spectroscopic data and consistent with a
 previous phenomenological IBM fit \cite{Samb82}. The theoretical method used in this work is robust, it is capable of
 correctly modeling the coexistence of shapes, hence, allows to gain a
 universal description of nuclear shapes, and will be applied to other
 heavy exotic nuclei in the future.

We thank the Tandem accelerator staff at the Wright Nuclear Structure
Laboratory, Yale University for their help during the experiment. This
work is supported by U.S. DOE under Grant
No. DE-FG02-91ER-40609. K.N. acknowledges the support through the JSPS
Postdoctoral Fellowships for Research Abroad. P.P. is indebted to the
Bulgarian Science Fund for the financial support under contract DFNI-E
01/2.

 \end{document}